\documentclass[10 pt]{article}
\usepackage{amsmath} % assumes amsmath package installed
\usepackage{amssymb}  % assumes amsmath package installed
\usepackage{makecell}
\usepackage{float}
\usepackage{graphicx}
\usepackage{caption}
\usepackage{subcaption}
\usepackage{spconf}

\usepackage{color}

\let\oldbibliography\thebibliography
\renewcommand{\thebibliography}[1]{%
  \oldbibliography{#1}%
  \setlength{\itemsep}{0pt}%
}
\title{
    Correlating Cellular Features with Gene Expression using CCA
}

\name{Vaishnavi Subramanian$^{\star}$ \ Benjamin Chidester$^{\dagger}$ \  Jian Ma$^{\dagger}$ \  Minh N. Do$^{\star}$ }

\address{$^{\star}$ Electrical and Computer Engineering, University of Illinois at Urbana-Champaign, USA\\
    $^{\dagger}$ Computational Biology, School of Computer Science, Carnegie Mellon University, USA}

\begin{document}

\maketitle
\thispagestyle{empty}
\pagestyle{empty}

%%%%%%%%%%%%%%%%%%%%%%%%%%%%%%%%%%%%%%%%%%%%%%%%%%%%%%%%%%%%%%%%%%%%%%%%%%%%%%%%
\begin{abstract}

%Complex diseases, such as cancer, have traditionally been studied using genetic data, or images alone. 
To understand the biology of cancer, joint analysis of multiple data modalities, including imaging and genomics, is crucial.
%The involved nature of gene-microenvironment interactions necessitates the use of algorithms which treat both data types equally.
We propose the use of canonical correlation analysis (CCA) and a sparse variant as a preliminary discovery tool for identifying connections across modalities, specifically between gene expression and features describing cell and nucleus shape, texture, and stain intensity in histopathological images.
Applied to 615 breast cancer samples from The Cancer Genome Atlas, CCA revealed significant correlation
%of 0.736 ($p \approx 1e^{-14}$) and 0.471, ($p \approx 7e^{-3}$) for CCA and Sparse CCA, respectively, 
of several image features with expression of PAM50 genes, known to be linked to outcome, while Sparse CCA revealed associations with enrichment of pathways implicated in cancer without leveraging prior biological understanding.
These findings affirm the utility of CCA for joint phenotype-genotype analysis of cancer.

\end{abstract}

%%%%%%%%%%%%%%%%%%%%%%%%%%%%%%%%%%%%%%%%%%%%%%%%%%%%%%%%%%%%%%%%%%%%%%%%%%%%%%%%

\section{Introduction}
Cancer is a complex disease arising from molecular alterations that interact with and obstruct normal biological processes and produce phenotypic changes.
Imaging modalities, such as microscopic imaging of hematoxylin-and-eosin (H\&E) stained slides %and radiology, 
and high-throughput genomics provide complementary information about the phenotypic traits (such as cell morphology) and molecular traits (such as gene expression and mutations) in a tumor.
The Cancer Genome Atlas (TCGA)~\cite{Koboldt2012} %and the Parkinson's Progression Markers Initiative (PPMI)~\cite{marek2011parkinson} -- JM: no need to mention Parkinson since it's irrelevant here.
provides rich resources of imaging, genomic, and clinical data and exemplifies the growing interest in comprehensive phenotypic and genomic data sets for disease understanding.
%To infer salient connections between the multiple modalities of such data sets, computational and statistical tools are needed.
%Therefore, computational methods that can connect different modalities are important to achieve a better understanding of diseases , resulting in the current area of imaging-genetics~\cite{bigos2010imaging}.

To explore connections between multiple data modalities, correlation analysis is the most straightforward approach.
However, since gene expression is the product of complex interacting cellular processes, cross-modality connections should be made that consider their relative levels and not just pairwise relationships.
A recent effort to explore correlation analysis for paired image and genomic data is the work by Cooper \textit{et al.}~\cite{Cooper2015}. The authors extracted human-annotated measures of necrosis and angiogenesis, along with cellular features, from histopathological images of glioblastomas and studied their correlation with gene expression.
To incorporate interactions between genes, each patient's gene expression was represented as a mixture of clustered gene signatures derived from the data.
%[Ben: will look for a reference for MRI]

Canonical correlation analysis (CCA)~\cite{hotelling1936relations} is an alternative approach for exploration that extends pairwise correlation analysis by considering linear combinations of the variables of each modality.
An advantage of CCA is that it requires no preprocessing of gene expression or image features to incorporate their interactions, but rather learns them via the linear model.
%Though the linear model is simple, it has been shown to be a powerful inference tool for analyzing other multimodal data, such as images paired with text~\cite{Hardoon2004}.
A sparsity-based extension is Sparse CCA (SCCA)~\cite{witten2009penalized}, which makes CCA possible for high-dimensional data with few samples and is particularly suited for gene expression.
%While these methods generally work with an independent set of variables, and a dependent set, the gene-microenvironment interaction is more subtle, and would benefit from methods that weigh both factors equally.

%Machine learning techniques such as multimodal Boltzmann machines \cite{srivastava2012multimodal} and bimodal deep autoencoders \cite{ngiam2011multimodal} allow for the equal importance of bimodal features. However, these techniques, based on deep learning, can be highly susceptible to overfitting on high-dimensional, low-sample data and will not generalize well. 
%

%To this end, we propose the use of canonical correlation analysis (CCA) \cite{hotelling1936relations} and its variants \cite{witten2009penalized}. %, to identify interesting relationships.
%CCA can discover correlations of mixtures of features from the two domains of interest, as compared to normal correlation analysis, which only considers pairwise correlations between individual features. 
%While allowing for more expressive associations between imaging and genetics, CCA is still simpler than deep learning approaches, and so the discovered associations will potentially generalize to other patients.

We explored the use of CCA 
% to affirm the utility of PAM50 genes \HL{in capturing the variation of molecular subtypes within breast cancer}, and the use of SCCA 
and SCCA
for unbiased discovery of connections between histopathological image features and gene expression of breast cancer tumors.
In particular, we extracted cellular features of shape, color, and texture from images using CellProfiler~\cite{Carpenter2006} and a reliable, efficient patch-based approach for nuclear segmentation using convolutional neural networks (CNNs)~\cite{Chidester2018}.
Using CCA, we discovered a significant correlation of 0.763 ($p \approx 1e^{-14}$) between the texture and shape features of cells and the expression of PAM50 genes, and enabled a separation of patients based on subtypes without leveraging specific subtype information. 
Using SCCA, we discovered a correlation of 0.471 ($p \approx 7e^{-3} $) between a subset of image features and genes. % involved in cell-related and immune-related pathways.
Pathway analysis of the selected subset of genes using DAVID~\cite{Huang2008} revealed a meaningful connection between cell size and several genes related to immune response.
Based upon these findings, we propose the use of CCA and its sparse variant as a preliminary discovery tool for imaging-genomic connections.

\begin{figure}[t]
        \centering
        \includegraphics[width=0.45\textwidth]{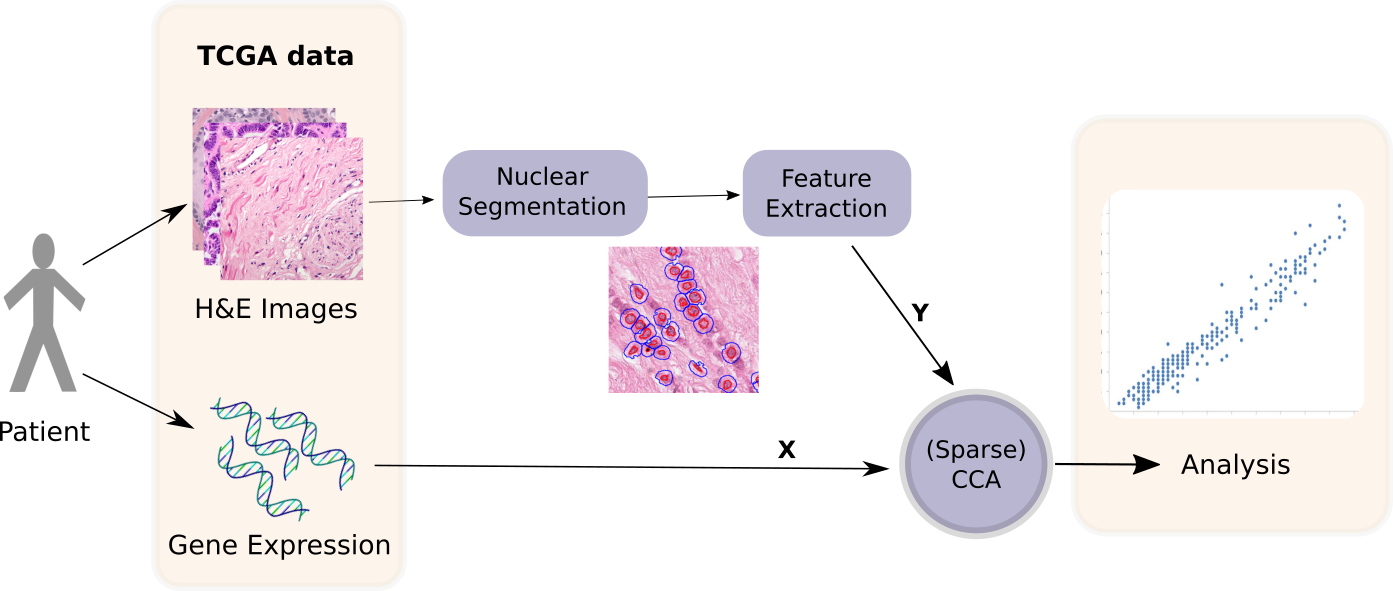}
	\caption{CCA workflow for imaging-genomics}
    \label{workflow}
    \vspace{-1.25em}
 \end{figure}

\section{Imaging-Genomics and CCA} % Canonical Correlation Analysis}

Our overall CCA workflow on paired histopathological images and gene expression of patients is shown in Fig~\ref{workflow}, which consists of nuclear and cellular segmentation and feature extraction, % patient feature summarization, 
and CCA and SCCA to discover significant connections between images and gene expression.
%For our analysis, we first extract cellular features from histopathological H\&E images of .
%We then summarized these cellular features by statistics of their distribution for each patient and applied CCA to a subset of the image features and gene expressions and Sparse-CCA to the full set of both.

\subsection{Nuclear Segmentation and Feature Extraction}  \label{cell_features}
 
In diagnosing breast cancer, the morphology and granularity of nuclei is an important indicator.
Therefore, we employed computational image analysis methods to extract quantitative features describing these qualities of cells and their nuclei.
%Estimating accurate nuclear and cellular boundaries through image analysis is critical for extracting reliable features and subsequent analysis.
We have developed a reliable and efficient patch-based CNN approach for segmentation~\cite{Chidester2018}, similar to that recently proposed by Janowczyk and Madabhushi~\cite{Janowczyk2016}, which scans an image patch-by-patch and produces a binary label for each patch, indicating if the center pixel of the patch is contained within a nucleus or not.
%We trained our CNN with data from~\cite{Janowczyk2016} along with our own manually-collected and labeled data from TCGA-BRCA whole slide images (WSIs).

Our patch-based CNN produces a full binary nuclear segmentation mask, which is then fed to CellProfiler~\cite{Carpenter2006}, along with the corresponding H\&E image, to
refine the segmentation % by smoothing boundaries and separating clumped nuclei, it 
and extract quantitative features describing the shape, texture, and color of the nuclei and cells.
To summarize these features across all of the cells of the image, the mean, standard deviation, and percentiles at increments of 10\% of the distribution of each feature are calculated.
This yielded $\sim 2400$ unique statistics of image features, which defined the image feature vector for the corresponding patient. %When only a subset of features is needed, we only capture the mean and standard deviation of the different shape, texture and color features.
Since analyzing entire WSIs is computationally demanding, we manually selected several representative patches from the tumor regions of each WSI and calculated the image feature vectors based on these patches, for each patient. % of the distribution of the extracted cellular features, which became the image feature vector for each patient. 

\subsection{Canonical Correlation Analysis}

For the imaging-genomics formulation, let $\mathbf{X}\in \mathbb{R}^{n \times p}$ denote the matrix of expression levels of $p$ genes for $n$ patients and $\mathbf{Y}\in \mathbb{R}^{n \times q}$ denote the matrix of $q$ image features for the same $n$ patients. 
To understand the information shared between $\mathbf{X}$ and $\mathbf{Y}$, we make use of CCA and its sparse variant.

Introduced by Hotelling~\cite{hotelling1936relations}, CCA is a method for determining the linear relationship between two sets of variables. 
Given two sets of variables, $\mathbf{X}$ and $\mathbf{Y}$, attributed to the same $n$ samples, CCA seeks linear combinations of the variables in each domain that are maximally correlated with each other.
Formally, CCA seeks $\mathbf{\alpha} \in \mathbb{R}^p$ and $\mathbf{\beta} \in \mathbb{R}^q$ that maximize the objective function
$$\max_{\mathbf{\alpha}, \mathbf{\beta}} \mathbf{\alpha}^T \mathbf{X}^T \mathbf{Y}\mathbf{\beta} \ \ \text{such that} \ \ \mathbf{\alpha}^T \mathbf{X}^T \mathbf{X\alpha} = \mathbf{\beta}^T \mathbf{Y}^T \mathbf{Y \beta} = 1,$$ where the columns of $\mathbf{X}$ and $\mathbf{Y}$ are standardized to mean zero and unit variance. 
The vectors $\mathbf{\alpha}$ and $\mathbf{\beta}$ are referred to as the canonical weights and $\mathbf{X\alpha}$ and $\mathbf{Y\beta}$ are the canonical variates.
This process can be repeated to find $k$ dimensions of canonical variates. Similar to principal component analysis, orthogonality constraints are imposed such that corresponding variates are orthogonal to previously found pairs.
The correlations of each variable of each domain with its corresponding canonical variate are called the \textit{canonical loadings}. 
For example, for image feature $f_1 $ and the first variate $\mathbf{Y}\beta_1$, both $\in  \mathbb{R}^p$, the loading $L(f_1, \mathbf{Y}\beta_1) =~\text{corr}(f_1, \mathbf{Y}\beta_1)$ where $\text{corr}(\cdot)$ is the Pearson's correlation. 
Here, we employ CCA to obtain the canonical weights, and hence the canonical variates, and look to identify the genes and image features of most importance in the variate space. 

For most of the genomic data used today, $n~\ll~\max(p,q)$, while CCA is only suitable when $n~\geq~\max(p,q)$. 
Applying CCA to high-dimensional, low-sample data therefore requires selecting a subset of the features in advance, or first mapping the features to a lower dimensional space, limiting the utility of the approach.
To overcome this issue, many versions of \textit{penalized CCA} have been proposed, which can work for high-dimensional data, while preserving interpretability.

We work with the formulation described by Witten \textit{et al.}~\cite{witten2009penalized}.
Called \textit{SCCA}, this method optimizes the objective function 
$$\max_{\mathbf{\alpha}, \mathbf{\beta}} \mathbf{\alpha}^T \mathbf{X}^T \mathbf{Y}\mathbf{\beta},$$ $$\text{such that} \ \| \alpha \|^2 \leq 1, \| \beta \|^2   \leq 1, P_x(\alpha)\leq c_x, P_y(\beta) \leq c_y,$$ 
where $P_x$ and $P_y$ are convex penalty functions, often chosen to impose sparsity.
For our analysis, we chose the $L_1$ penalty function. 
For multiple variates, the algorithm is iterated. %an iterative algorithm is employed. % Unlike CCA, explicit orthogonality between successive variates is not enforced.
%Sparse CCA enables us to work with the entire set of image features and genes, with the possibility of revealing new connections.

\section{Results and Discussion}

\begin{figure*}[t]
\vspace{-1.5em}
\centering
 \hspace*{0cm}
     \begin{subfigure}[c]{0.45\textwidth}
        \centering
         \hspace{-0.5cm}
         \includegraphics[width=\textwidth]{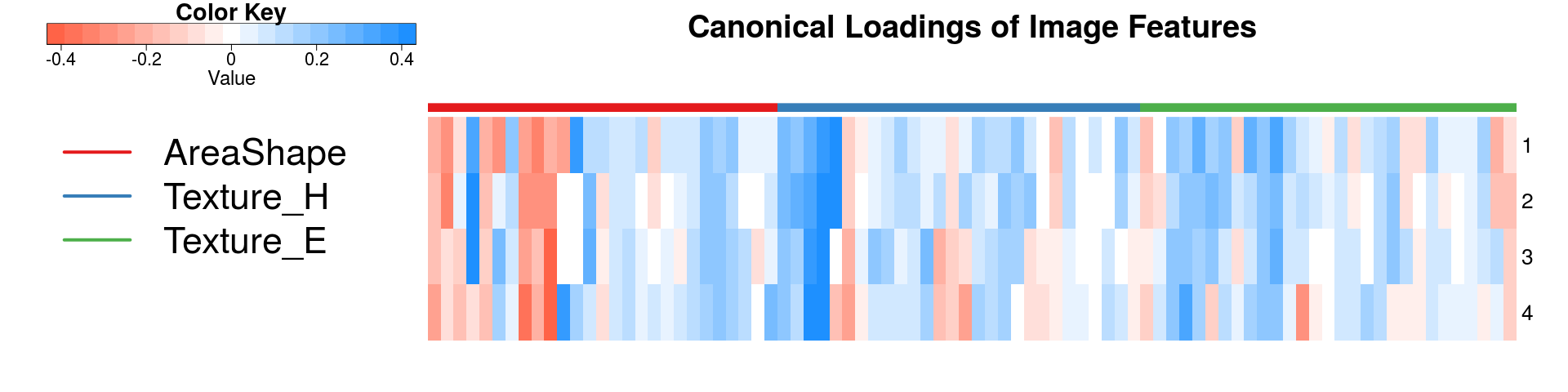}
      \hspace{-0.5cm}\caption{ }
    \end{subfigure}%
   	~
     \begin{subfigure}[c]{0.18\textwidth}
        \centering
%         \hspace{-1cm}
       \includegraphics[trim={7cm 0 0 0},clip,width=\textwidth]{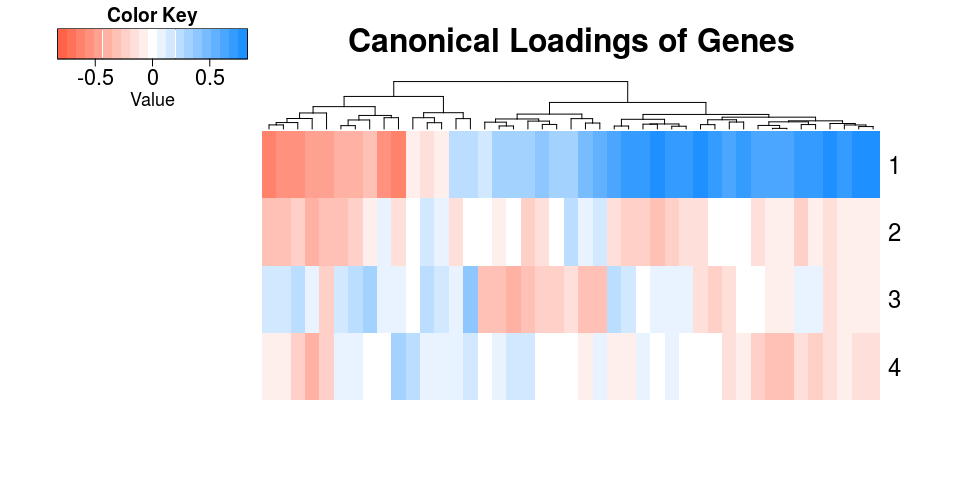}
      \caption{ }
    \end{subfigure}%
    ~
	\begin{subfigure}[c]{0.37\textwidth}
        \centering
%          \hspace*{-1cm}
        \includegraphics[width=\textwidth]{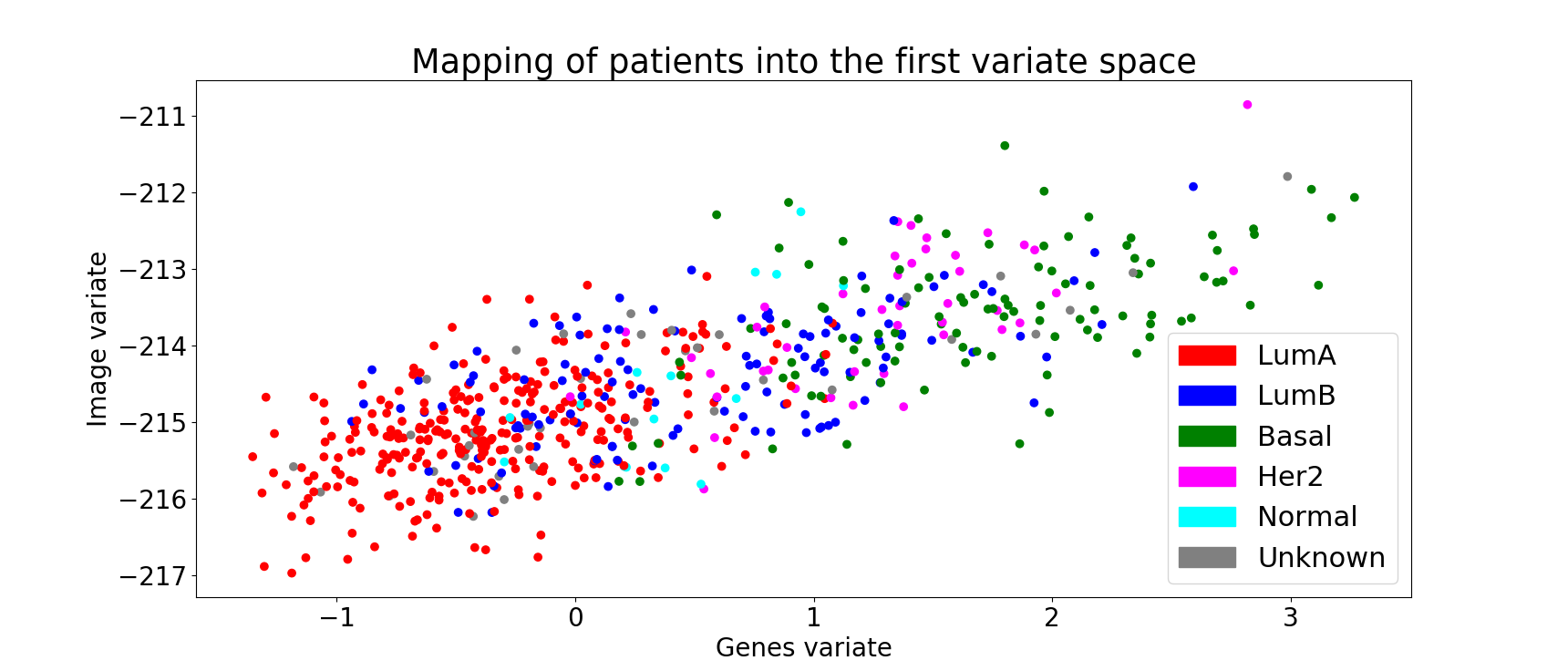}
        \caption{ }
    \end{subfigure}%
    
%    \caption{Canonical loadings of image features (left) and expression of genes (right) based on CCA (horizontal axis: genes/image-features, vertical axis: variate number).}
%     \label{fig:heatmap}
% \end{figure*}

% \begin{figure*}[t!]
% \centering
 \vspace*{0.5em}
    \begin{subfigure}[t]{0.58\textwidth}
        \centering
         \hspace*{-0.5cm}
       \includegraphics[width=\textwidth]{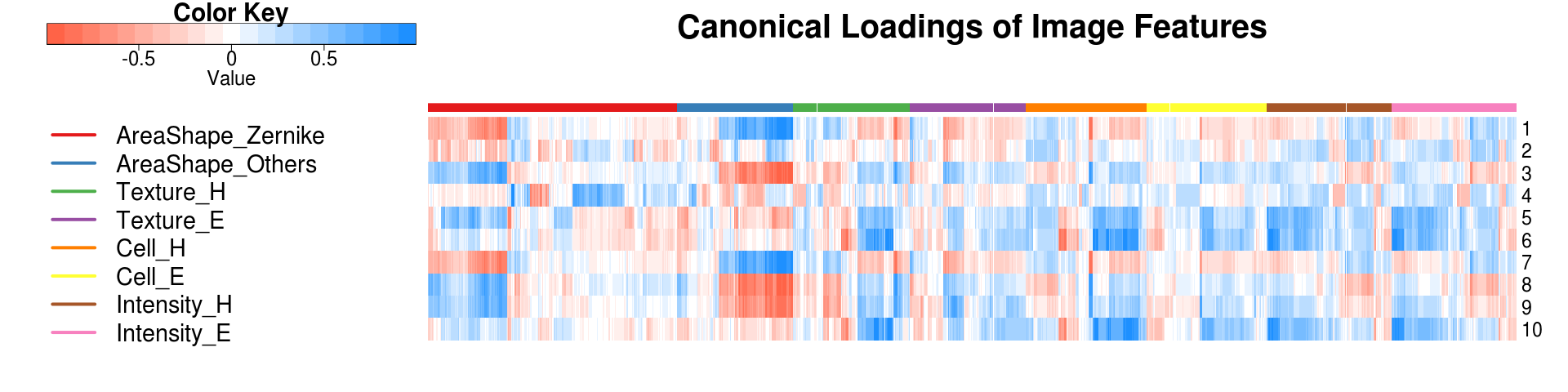}
     \caption{ }
    \end{subfigure}%
     ~
\begin{subfigure}[t]{0.42\textwidth}
        \centering
         \hspace*{-0.5cm}
        \includegraphics[trim={13.55cm 0 0 0},clip, width=\textwidth]{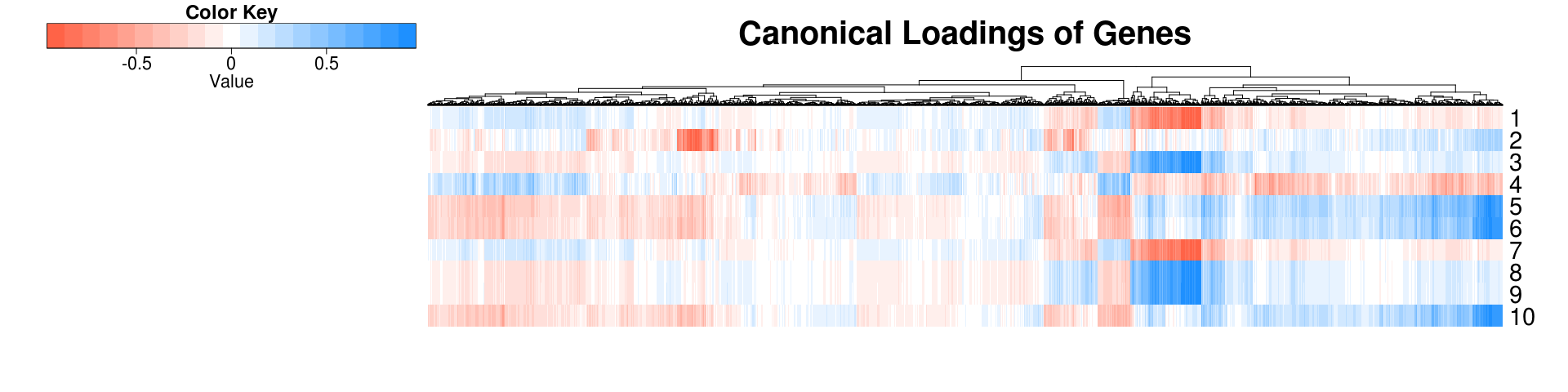}
         \hspace*{-0.5cm}
         \caption{ }
    \end{subfigure}%

\caption{Canonical loadings of image features (a)(d) and expression of genes (b)(e) based on CCA(top row) and SCCA(bottom row), horizontal axis:~genes/image-features, vertical axis: variate number, (c) shows the mapping of patients onto the 1st variate}

    \vspace{-1.0em}
    \label{fig:heatmap}
\end{figure*}

% \begin{figure*}[t!]
% \centering
%  \hspace*{0cm}
    
%      \includegraphics[width=0.9\textwidth]{images/heatmap_image}
%      \includegraphics[width=0.9\textwidth]{images/heatmap_genes}
%     \caption{Correlation of image features (top) and expression of genes (bottom) with the canonical variates learned by Sparse CCA (horizontal axis: genes/image-features, vertical axis: variate number)}
%     \label{fig:heatmap_sparse}
% \end{figure*}

% \begin{figure}[t!]
% \centering
%  %\hspace*{-1.2cm}
%     \begin{subfigure}[t]{0.25\textwidth}
%         \centering
%          \hspace*{-0.5cm}
%         \includegraphics[width=\textwidth]{images/CCA_feature_embeddings}
%         \caption{CCA loading}
%     \end{subfigure}%
%      ~
%     % \hspace*{-0.45cm}
%      \begin{subfigure}[t]{0.25\textwidth}
%         % \hspace*{-1.2cm}
%          \centering
%         \includegraphics[width=\textwidth]{images/sparse_CCA_feature_embeddings}
%         \caption{Sparse CCA loading }
%     \end{subfigure} %

%     \caption{Canonical loadings of CCA and Sparse CCA (darker shade in (b) implies feature had non-zero weight in either of first 2 dimensions)}
%     \label{embedding}
% \end{figure}

We applied the overall method %of nuclear and cellular segmentation and feature extraction, and CCA of cell features and gene expression, 
on 615 breast invasive carcinoma (BRCA) patients from TCGA.
Histopathological images for TCGA patients are in the form of whole slides (WSIs), and in order to reduce the computational burden of image analysis and to avoid contamination in the analysis by normal cells near the tumor, we manually selected up to fifteen representative patches of 1000$\times$1000 pixels from each WSI in the tumor region for segmentation and feature extraction.
Gene expression was retrieved from TCGA using cBioPortal, which normalized expression levels to z-scores. The analyses are done in R using the default CCA package and the SCCA package provided by Witten \textit{et al.}~\cite{witten2009penalized}.

\subsection{Using CCA}

%% FIGURE

% \begin{figure}[t!]
%   \hspace*{-1cm}
%         \centering
%         \includegraphics[width=0.35\textwidth]{images/data_correlation_p_value}
% 	\caption{Variation of p-value and correlation with dimension of hidden variable for CCA}
%     \label{data_corr_pval}
%  \end{figure}

%% FIGURE

%\begin{figure*}[!ht]
%\centering
%    \begin{subfigure}[c]{0.35\textwidth}
%        \centering
%      %   \hspace{-1.8cm}
%         \includegraphics[width=\textwidth]{images/gene_features.png}
%      % \caption{ }
%    \end{subfigure}%
%   	~
%     \begin{subfigure}[c]{0.55\textwidth}
%        \centering
%        %\vspace{0.01em}
%       % \hspace{-0.4in}
%        \includegraphics[width=\textwidth]{images/sparse_gene_features.png}
%       % \caption{ }
%    \end{subfigure}%
%   
%   \caption{Top loadings of genes and image features with 1st and 2nd variates of CCA (left), and Sparse CCA (genes)}
%   \label{cca_loadings}
%\end{figure*}

% \begin{figure*}[t]
% \centering
%  \hspace*{0cm}
    
To apply CCA to the extracted image features and gene expression of TCGA-BRCA patients, we first had to select a smaller subset of both image features and genes such that $\max(p,q)< n$.
Of the image features, we used only the mean and standard deviation of the shape, texture and color features, which resulted in 84 image features per patient. 
As a meaningful subset of genes to analyze, we chose the PAM50 set of 50 genes, which has been shown to be discriminative of the general grouping of patients into molecular subtypes~\cite{parker2009supervised}. %The set of TCGA-BRCA patients we analyzed comprised all molecular subtypes.

Using CCA on these restricted sets of variables, we found four canonical variates of statistical significance (p-value less than 0.05, computed using Wilk's lambda statistic) with strong correlation ($\left\{0.76, 0.64, 0.61, 0.59 \right\}$ respectively).
Beyond the first four variates, the significance of the correlation quickly dropped.
%, implying that the variance of the correlation about the variate is high and the null hypothesis of no correlation between the variates cannot be rejected.
%The standard approach to interpreting the canonical variates is to look at the weights (sign and magnitude) of the weights.
%While the canonical weights define the directions which are meaningful for correlating the two datasets, the canonical loadings enable the identification of features most correlated with these directions. These features can potentially assign meaning to the corresponding variate. 
%However, interpreting these can be challenging, e.g., a feature could have a low weight either because it is irrelevant to the covariate, or because it has been shadowed out of the relationship because of a high degree of collinearity with a collection of other features. However, the \textit{canonical loadings} of variables of each domain with their respective canonical variates provide insight into their interpretation.
%A variable that is highly correlated with a canonical variate is well-explained by that canonical variate.
To interpret the learned canonical variates, we examined the canonical loadings of each image feature and gene with each variate, which are shown in Fig.~\ref{fig:heatmap}(a)(b).

We observe that the first canonical variate is highly correlated with many PAM50 genes, with correlations as high as $0.8$, which implies that this variate is highly representative of PAM50 expression.
%This high correlation implies that the first variate is able to capture the variance of a significant number of genes.
The loadings of the image features in Fig.~\ref{fig:heatmap}(a) are grouped by category, which reveals the strongest correlation for most variates is with several texture features of the hematoxylin stain, area, and shape.
The first variate shows a strong positive correlation particularly with texture features describing the entropy and variance of the hematoxylin stain within the nucleus and shape features describing the nucleus.  Subsequent variates showed much lower loadings, so while still significantly correlated within their imaging counterpart, the interpretation is not as clear.
% Shape features, in particular the form factor of the nucleus, also showed strong positive correlation.

To further understand the first variate, the $615$ patients are mapped into the corresponding variate space. The scatter plot of the mappings ($\alpha^T\mathbf{X}$,  $\beta^T\mathbf{Y}$ on $x$ and $y$ axes, respectively) is shown in Fig.~\ref{fig:heatmap}(c), with the color representing the true subtype. Luminal~A patients are clustered towards the left, and Basal patients to the right, while HER2 and Luminal~B patients are spread out in between. This spread of the subtypes is, interestingly, in accordance with the expected prognosis of the patients. It is also noted that the range of values in the image variate is considerably smaller than those of the genes, suggesting that we should consider a more diverse set of image features.

Though CCA can potentially pick any relevant linear combination of features resulting in any possible ordering of the subtypes, the result of information from both modalities was a meaningful order of subtypes: Luminal~A, Luminal~B, HER2 and Basal. 
Thus, while it is known that the PAM50 gene set is indicative of molecular subtype, CCA was able to identify the particular combination of genes and image features which can map patients into the subtype, without leveraging particular subtype information.

\begin{figure*}[t]
\vspace{-1.0cm}
\centering
 \hspace*{0cm}
        \vspace{-0.5em}
     \includegraphics[width=\textwidth]{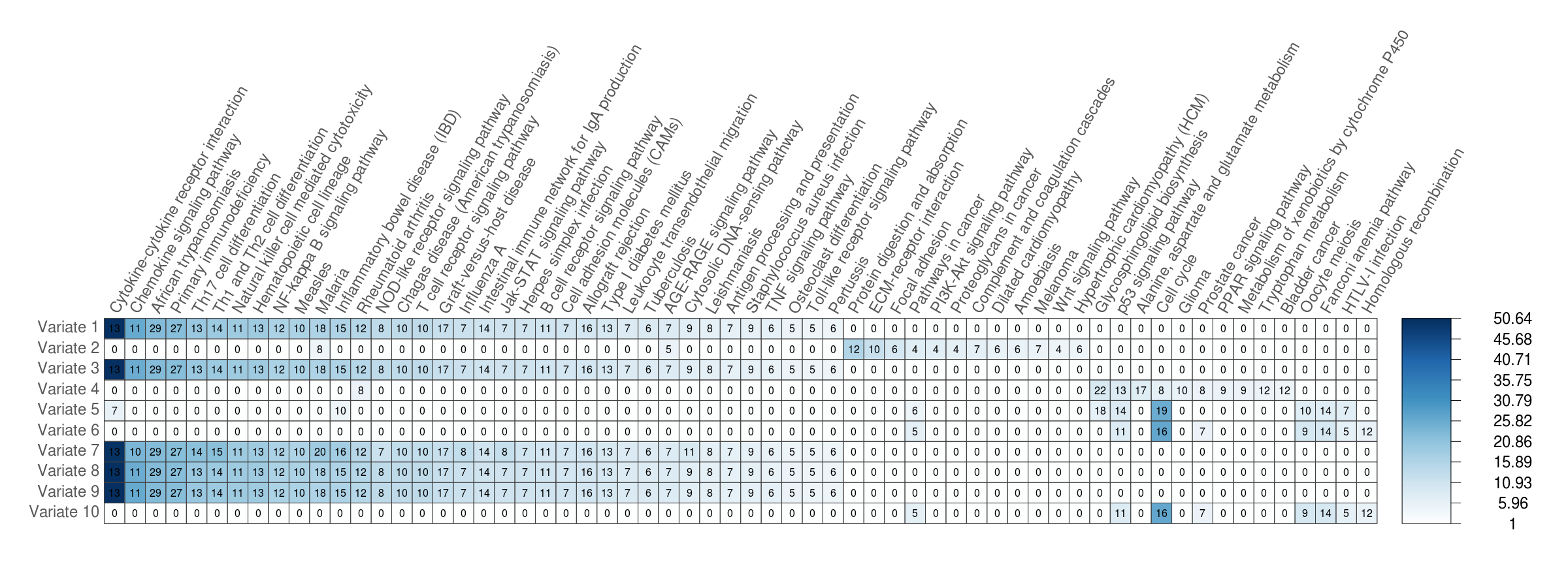}
    \vspace{-1.5em}
\caption{Plot of variates vs pathways defined by genes with correlation~$>$~0.35 based on SCCA (Intensity of color represents the -$\log $(p-value), the number is the percentage of pathway genes overlapping with $0$ meaning not computed).}
    \vspace{-1.0em}
    \label{fig:p_value_genes}
\end{figure*}

% \begin{figure}[t!]
%     \centering
%     \begin{subfigure}[t]{0.15\textwidth}
%         \centering
%         \includegraphics[height=2.2cm]{images/high_corr1_small}
%         \caption{}
%     \end{subfigure}%
%     ~ 
%     \begin{subfigure}[t]{0.15\textwidth}
%         \centering
%         \includegraphics[height=2.2cm]{images/low_corr1_small}
%         \caption{}
%     \end{subfigure}%
%     ~ 
%     \begin{subfigure}[t]{0.15\textwidth}
%         \centering
%         \includegraphics[trim={0.4cm 0 0 0},clip,height=2.2cm]{images/low_corr2_small}
%         \caption{}
%     \end{subfigure}%
%     \caption{Image samples of patients with high correlation (a) and low correlation (b)(c) with 1st variate}
%     \vspace{-1.5em}
%     \label{example_patients}
% \end{figure}

\subsection{Using SCCA}

In contrast to CCA,  we were able to analyze all image features and genes using SCCA, allowing the algorithm to discover which subset of each is most correlated.
Using an L1 penalty factor of 0.1 for both image and genomic variables, we obtained sets of 45-60 genes and 30-45 image features with non-zero weights for each of the ten canonical variates, respectively, with
correlations %of  $\{ 0.47, 0.37, 0.45, 0.46, 0.40, 0.35, 0.44, 0.46, 0.45, 0.36 \}$ 
in the range of 0.35-0.47, with an overall p-value of 0.001.
To interpret the learned canonical variates of SCCA, we make use of the loadings as before, as shown in Fig.~\ref{fig:heatmap}(d)(e). % shows the loadings of all image features and genes with the ten learned canonical variates. 
The category of `cell' indicates that the feature is of the cytoplasmic region surrounding the nucleus, which mostly describe area and shape.
All other features are extracted from the nucleus only.

%The correlation plot for gene expressions and the canonical variates reveals a grouping of about 500 genes which have a strong correlation with at least one of the variates.

%Referring to Fig.~\ref{cca_loadings} we see that the highest weighted genes with the first and second variates have values of correlation of $\approx 0.3$, ignoring signs. As earlier, this could imply that the variate is capturing an aggregate of various genes, rather than individual ones.

Since SCCA can consider all genes and image features, it can reveal novel, unbiased phenotype-genotype associations.
We selected  genes whose expression levels were highly correlated ($>~0.35$) with the canonical variates discovered by SCCA and investigated their collective function using the online functional annotation tool DAVID~\cite{Huang2008}, %and the ConsensusPathDB
which can test for association of gene sets with KEGG pathways.
The KEGG pathways significantly associated are shown in Fig~\ref{fig:p_value_genes}. 
%It is interesting to note that pathways directly associated with cancer were identified (including pathways in cancer, central carbon metabolism in cancer and viral carcinogenesis).

The first variate and others showed a similar correlation pattern with both image features and gene expressions, which is likely a result of the lack of enforcement of orthogonality by SCCA.
DAVID revealed that, for the first variate, the highly correlated genes were strongly associated with pathways related to immune response, including primary immunodeficiency, natural killer cell mediated cytotoxicity, and to lymphocytes, including Th1 and Th2 cell differentiation, T-cell and B-cell receptor signaling, and NF-kappa B signaling.
Fig.~\ref{fig:heatmap}(d)(e) shows that the expression of these genes has a strong %inverse
correlation with area and shape features through the latent canonical variates.
Given that lymphocytes are easily distinguished by their small size and circular shape, we could hypothesize that these canonical variates are capturing image and genomic descriptions of the presence of lymphocytes within the tumor, which is indeed a biologically relevant association for cancer. % To support this hypothesis, some sample images of patients with high and low correlations with the first variate are shown in Fig~\ref{example_patients}, where the difference in the presence of lymphocytes~(small, dark and dense regions) can be seen. 
%is involved in the control of several cell processes implicated in cancer, including adhesion, migration, differentiation, proliferation, and apoptosis. 

%While the shape features of cells can potentially capture stages of cell cycle, features capturing the spatial distribution of cells in the tissue might be better indicative of the cell processes. 
Variates five, six, and ten 
%are not indicative of area or shape, but instead 
capture texture and cell hematoxylin features, which are indicative of DNA content, and intensity features of both stains.
These variates were found to be correlated with gene sets associated with the cell cycle and p53 signaling pathways (related to DNA damage repair and apoptosis), %as well as the cell cycle, 
all of which have important implications for tumor development.
The second variate too could have implications for cancer, as it was associated with pathways involved in cell processes such as cell maintenance (ECM-receptor interaction), adhesion (focal adhesion), and proliferation (Wnt signaling and proteoglycans in cancer), and the cycle (PI3K-Akt signaling), though the lack of strong correlation with particular image features necessitates further investigation for a clear interpretation.
% The other variates 

% Cell-matrix adhesions play essential roles in important biological processes including cell motility, cell proliferation, cell differentiation, regulation of gene expression and cell survival. Focal adhesion pathway 

% The feature embeddings based on loadings of all features for Sparse CCA (Fig.~\ref{embedding}(b) indicates equal spread of features along both dimensions. Fig.~\ref{embedding}) focuses on those features which received a non-zero weight in the first two variates. The highly dispersed nature of the features indicates that most of the selected features are very well correlated (correlation of $\approx 1 $, ignoring sign) with the first and second variates, implying a strong correlation directly between the selected features. 

% \HL{- Explain Figure 2(c) better - Results discussion more relevant and conclusive}

\section{Conclusions and Future Work}
We have demonstrated the utility of CCA and SCCA in discovering connections between cellular features and gene expressions for breast cancer.
The learned canonical variates represent latent spaces that link the two modalities and provide insight into their joint variation.
Their biological relevance was shown through their association with diverse pathways with implications for cancer, and could benefit from a more diverse range of image features. % For SCCA, imposing orthogonality in the variates and understanding the sensitivity of the penalty factor for SCCA would be important for use in a clinical setting.
We envision that such a correlation analysis could be a preliminary step in studies of phenotypic and genomic traits, with follow-up affirmation and testing by pathologists and biologists, toward new insights into genetic diseases.

% The first variate and others showed a similar correlation pattern with both image features and gene expressions, which is likely a result of the lack of enforcement of orthogonality by SCCA.

%Such connections are highly informative for the subsequent integrative analysis of imaging data and genomic data from cancer samples.
%We believe our proposed CCA and Sparse CCA approaches would be useful preliminary discovery tool for imaging-genomics analysis. 

\bibliographystyle{IEEEbib}
{\fontsize{9pt}{10pt}\selectfont\bibliography{bhi2018,vaishnavi}}
% \bibliography{bhi2018,vaishnavi}
\end{document}